\documentclass{ws-mpla}
\usepackage[super]{cite}
\usepackage{graphicx}
\begin{document}

\markboth{Z. Chang, X. Li, H.-N. Lin, S. Wang}{Constraining the anisotropy of the universe from SNe and GRBs}


\title{Constraining the Anisotropy of the Universe from Supernovae and Gamma-ray Bursts}

\author{\footnotesize Zhe Chang}
\address{Institute of High Energy Physics, Chinese Academy of Sciences, 100049 Beijing, China\\
Theoretical Physics Center for Science Facilities, Chinese Academy of Sciences, 100049 Beijing, China}

\author{\footnotesize Xin Li}
\address{Institute of High Energy Physics, Chinese Academy of Sciences, 100049 Beijing, China\\
Theoretical Physics Center for Science Facilities, Chinese Academy of Sciences, 100049 Beijing, China}

\author{\footnotesize Hai-Nan Lin}
\address{Institute of High Energy Physics, Chinese Academy of Sciences, 100049 Beijing, China\\
linhn@ihep.ac.cn}

\author{\footnotesize Sai Wang}
\address{Institute of High Energy Physics, Chinese Academy of Sciences, 100049 Beijing, China\\}

\maketitle
\pub{Received (Day Month Year)}{Revised (Day Month Year)}

\begin{abstract}
Recently, an anisotropic cosmological model was proposed. An arbitrary 1-form, which picks out a privileged axis in the universe, was added to the Friedmann-Robertson-Walker line element. The distance-redshift relation was modified such that it is direction dependent. In this paper, we use the Union2 dataset and 59 high-redshift gamma-ray bursts to give constraints on the anisotropy of the universe. The results show that the magnitude of anisotropy is about $D=-0.044\pm0.018$, and the privileged axis points towards the direction $(l_0,b_0)=(306.1^{\circ}\pm 18.7^{\circ},-18.2^{\circ}\pm 11.2^{\circ})$ in the galactic coordinate system. The anisotropy is small and the isotropic cosmological model is an excellent approximation.
\end{abstract}
\keywords{cosmology, anisotropy, supernovae, gamma-ray bursts}

\section{Introduction}	

The ${\rm \Lambda}$CDM model is widely accepted as the standard model in modern cosmology. It is based on the cosmological principle, which says that the universe is homogeneous and isotropic at large scales. The ${\rm \Lambda}$CDM model is well consistent with current observations on the cosmic microwave background (CMB) radiation \cite{Spergel:2007hy,Komatsu:2011fb}, the large scale structure of the cosmos \cite{Eisenstein:2005su}, the statistics of galaxies \cite{TrujilloGomez:2011yh}, the halo power spectrum \cite{Reid:2010xm}, and so on. The tiny fluctuations of the CMB temperature show that the cosmological principle is an excellent assumption \cite{Smith:2009jr}.

In spite of the successes mentioned above, the ${\rm \Lambda}$CDM model also faces many challenges. It was found that there exists a large bulk-flow velocity at scales up to about $100h^{-1}$ Mpc \cite{Kashlinsky:2009ut,Watkins:2009hf,Lavaux:2010th}. However, the ${\rm \Lambda}$CDM model predicts a much smaller velocity at a much smaller scale. The alignments of low multipoles in the CMB angular power spectrum seem to point towards certain directions \cite{Lineweaver:1996xa,Tegmark:2003,Bielewicz:2004en,Frommert:2009qw,Copi:2010,Chang:2013lxa}, which also conflicts with the prediction of ${\rm \Lambda}$CDM model. The large-scale alignments of the quasar polarization vectors are not randomly oriented over the sky with a high probability \cite{Hutsemekers:2005iz,Hutsemekers:2008iv}. The alignment effect seems to be prominent along a particular axis. All of these anomalies, although not confirming, imply that the universe may be anisotropic and there may exist a privileged axis.

As the standard candle, the type Ia supernovae (SNe Ia) have been widely used to study the possible anisotropy of the universe. In the framework of ${\rm \Lambda}$CDM model, Antoniou \& Perivolaropoulos used the hemisphere comparison method to search for a preferred axis in the full Union2 dataset \cite{Antoniou2010}. They found that at a certain direction, the accelerating expansion rate of the universe reaches its maximum. Using the same data and method, Cai \& Tuo found a similar preferred axis in the context of $\omega$CDM model \cite{CaiTuo2012}. Kalus et al. studied the SNe Ia with low redshift $z<0.2$ and also found a direction at which the expansion rate reaches its maximum \cite{Kalus:2013zu}. Zhao et al. studied the anisotropy of cosmic acceleration by dividing the Union2 dataset into 12 subsets, and found a dipole effect whose direction is approximately perpendicular to the direction of the CMB kinematic dipole \cite{Zhao:2013yaa}. Although the magnitude of isotropy is small enough such that the ${\rm \Lambda}$CDM model is an excellent approximation, an anisotropic cosmological model cannot be ruled out.

Very recently, Chang et al. proposed an anisotropic cosmological model in the Randers spacetime \cite{Chang2013}. An arbitrary 1-form was added to the Friedmann-Robertson-Walker (FRW) line element. This 1-form picks out a preferred axis in the universe. The distance-redshift relation was modified such that it is direction-dependent. The constraint from the Union2 dataset gave the magnitude of anisotropy and the direction of the privileged axis. However, the redshifts of SNe Ia in the Union2 dataset are no more than 1.4. It is questionable whether the results can be extended to the high-redshift region. The gamma-ray bursts (GRBs) provides a good supplementary to SNe Ia. The furthest GRB detected at present has redshift as high as 8 or more. For example, the redshift of GRB 090424 is about 8.1 \cite{Salvaterra2009}. In fact, GRBs have already been used by some authors to study the Hubble diagram \cite{Schaefer2007,Wei2010,Cai:2013lja,Velten:2013ej,Wei:2013xx}.

In this paper, we combine the Union2 dataset with GRBs to constrain the parameters of the anisotropic cosmological model proposed by Chang et al.. The rest of the paper is organized as follows: In section \ref{sec:model}, we briefly review the anisotropic cosmological model in the context of Randers spacetime. In section \ref{sec:numerical}, we use the Union2 dataset and 59 long GRBs to constrain the parameters of the model. We find that there is obviously a privileged axis in the universe, and the anisotropy of the universe is at the order of magnitude $D=-0.044\pm0.018$. Finally, discussions and conclusions are given in section \ref{sec:discussion}.

\section{The Anisotropic Cosmological model in the Randers spacetime}\label{sec:model}

The line element in the Randers spacetime can be written as the FRW line element added by an extra 1-form \cite{Chang2013}
\begin{equation}\label{eq:Randers line element}
  d\tau=\overline{d\tau}+\tilde{b}_{\mu}({\mathbf x})dx^{\mu},
\end{equation}
where
\begin{equation}\label{eq:FRW line element}
  \overline{d\tau}=\sqrt{dt^2-a^2(t)(dx^2+dy^2+dz^2)}
\end{equation}
is the FRW line element in the ${\rm \Lambda}$CDM model, and the 1-form $\tilde{b}_{\mu}({\mathbf x})dx^{\mu}$ picks out a privileged axis in the universe. ${\mathbf x}=(t,x,y,z)$ is the four-dimensional spacetime coordinate. Chang et al. have showed that the anisotropy of Hubble diagram stems from the spatial components of the privileged axis $\tilde{b}_{\mu}({\mathbf x})$. Without loss of generality, one can choose the Cartesian coordinate system (CCS) such that the z-axis is towards the privileged direction. Furthermore, assuming that the 1-form depends only on the time coordinate, then Eq.(\ref{eq:Randers line element}) simplifies to
\begin{equation}
  d\tau=\sqrt{dt^2-a^2(t)(dx^2+dy^2+dz^2)}+\tilde{b}_3(t)dz.
\end{equation}

Following the similar procedure of general relativity, one can calculate the connections and curvatures from the line element. Substituting the connections and curvatures into the Einstein equations, we get the modified Friedmann equation in the Randers spacetime. It reads
\begin{equation}\label{eq:modified Friedmann equation}
  3\left(\frac{\dot{a}}{a}\right)^2+\frac{1}{a^4}\left[(\dot{a}^2-2a\ddot{a})\tilde{b}_3^2-a\dot{a}(\tilde{b}_3^2)^.\right]=8\pi G\rho,
\end{equation}
where the dots denote the temporal derivative $d/dt$, $G$ is the gravitational constant and $\rho$ is the energy density of cosmic inventory. When $\tilde{b}_3\equiv 0$, Eq.(\ref{eq:modified Friedmann equation}) reduces to the well-known Friedmann equation.

In the anisotropic cosmological model, the redshift of an object locating at the direction $\hat{{\bf p}}$ and time $t$ can be modified as
\begin{equation}\label{eq:modified redshift}
  1+z(t,\hat{\bf p})=\frac{1}{a(t)}[1-D(\hat{\bf n}\cdot\hat{\bf p})],
\end{equation}
where $\mid D \mid \ll 1$ represents the magnitude of anisotropy of the universe and $\hat{\bf n}$ is the direction of the privileged axis. When deriving Eq.(\ref{eq:modified redshift}), we have already set the scale factor today to be unity, ie., $a(t_0)=1$. Here and after, a hat over a vector denotes the unit vector of that direction. In the special CCS where the $z$-axis is exactly towards the privileged direction, Eq.(\ref{eq:modified redshift}) reduces to
\begin{equation}\label{eq:modified redshift2}
  1+z(t,\cos\theta)=\frac{1}{a(t)}[1-D\cos\theta],
\end{equation}
where $\theta$ is the angle between $\hat{\bf n}$ and $\hat{\bf p}$.

Since the universe is anisotropic, the luminosity distance-redshift relation is orientation-dependent. It can be rewritten as \cite{Chang2013}
\begin{equation}\label{eq:distance-redshift}
  d_L=(1+z)\frac{c}{H_0}\int_0^z \frac{(1-D\cos\theta)^{-1}dz}{\sqrt{\Omega_M\left(\frac{1-D\cos\theta}{1+z}\right)^{-3}+\Omega_{\Lambda}}},
\end{equation}
where $c$ is the speed of light in vacuum, and $H_0$ is the Hubble constant. When the anisotropy vanishes, i.e., $D=0$, Eq.(\ref{eq:distance-redshift}) returns back to that of ${\rm \Lambda}$CDM model. It is more convenient to transform the luminosity distance to the dimensionless distance modulus, which is defied as
\begin{equation}\label{eq:distance modulus}
  \mu\equiv5\log_{10}\frac{d_L}{{\rm Mpc}}+25.
\end{equation}

\section{Numerical Constraints from SNe Ia and GRBs}\label{sec:numerical}

In this section, we will use the Union2 compilation and GRBs to constrain the model parameters. Before proceeding, some coordinate systems should be clarified. Corresponding to the galactic coordinate system (GCS), we define a CCS, whose origin locates at the center of the GCS. The $z$-axis of the CCS is towards the north pole $(l,b)=(0^{\circ},90^{\circ})$, the $x$-axis is towards the point $(l,b)=(0^{\circ},0^{\circ})$, and the $xyz$ axes comprise the right-handed set. Thus, the orientation of an object with galactic coordinate $(l,b)$ can be written in the CCS as
\begin{equation}
  \hat{\bf p}=\cos(b)\cos(l)\hat{\bf i}+\cos(b)\sin(l)\hat{\bf j}+\sin(b)\hat{\bf k},
\end{equation}
where $\hat{\bf i}$, $\hat{\bf j}$ and $\hat{\bf k}$ are the unit vectors along the $x$, $y$ and $z$ axes, respectively. Furthermore, we suppose that the privileged axis is along the direction
\begin{equation}
  \hat{\bf n}=\cos(b_0)\cos(l_0)\hat{\bf i}+\cos(b_0)\sin(l_0)\hat{\bf j}+\sin(b_0)\hat{\bf k}.
\end{equation}
The cosine of the angle $\theta$ between $\hat{\bf p}$ and $\hat{\bf n}$ is given as $\cos\theta=\hat{\bf p}\cdot\hat{\bf n}$.

We first use the Union2 dataset \cite{Amanullah2010} to constrain the model parameters. The Union2 dataset consists of 557 SNe Ia, all of which have well observed reshift, distance modulus and orientation in the equatorial coordinate system (ECS). We define $\chi^2$ as
\begin{equation}
  \displaystyle\chi^2=\sum_{i=1}^N\left(\frac{\mu_{\rm th}^{(i)}-\mu_{\rm obs}^{(i)}}{\sigma_{\mu}^{(i)}}\right)^2,
\end{equation}
where $\mu_{\rm th}$ is the theoretical distance modulus calculated from Eqs.(\ref{eq:distance-redshift}) and (\ref{eq:distance modulus}), $\mu_{\rm obs}$ and $\sigma_{\mu}$ are the observed distance modulus and its uncertainty, respectively, and $N$ is the total number of SNe Ia. We perform the least-$\chi^2$ method to determine the parameters. The model have five parameters in total, ie., the magnitude of the anisotropy $D$, the orientation of the privileged axis $(l_0,b_0)$, the matter component $\Omega_M$ and the Hubble constant $H_0$. It should be noted that the observed orientations of the SNe Ia are given in the ECS. We should firstly transform them into the GCS. The detailed transformation between the two systems is excellently described in the book of Peter \cite{Peter1981}.

As the zeroth order approximation, namely $D\equiv0$, the Randers-type cosmology reduces to the ${\rm \Lambda}$CDM model. The best fit to the ${\rm \Lambda}$CDM model gives $(\Omega_M, h_0)=(0.270\pm 0.020, 0.700\pm 0.004)$. Where $H_0\equiv 100h_0~{\rm km~s}^{-1}{\rm Mpc}^{-1}$. Here and after, the errors quoted are of $1\sigma$. Next, we fix $(\Omega_M,h_0)$ to their central values and perform a three-parameter fit. The results show that the magnitude of the anisotropy is $D=-0.044\pm 0.019$, pointing towards the direction $(l_0,b_0)=(306.2^{\circ}\pm 19.3^{\circ}, -18.2^{\circ}\pm 11.6^{\circ})$. It should be mentioned that there exists another equivalent solution, which $D$ changes its sign and the privileged axis changes to its opposite direction, ie., $D\longrightarrow -D$ and $(l_0,b_0)\longrightarrow (l_0-180^{\circ},-b_0)$. This can be easily see from Eq.(\ref{eq:distance-redshift}), since $D$ is always multiplied by $\cos\theta$. For simplicity, we constrain $D$ to be negative. To see any possible dependence of the anisotropy on the values of $(\Omega_M,h_0)$, we set all parameters free and perform a five-parameter fit. The results are give as $(\Omega_M,h_0)=(0.268\pm 0.020, 0.703\pm 0.004)$, $D=-0.049\pm 0.019$, and $(l_0,b_0)=(307.0^{\circ}\pm 16.7^{\circ}, -14.5^{\circ}\pm 11.0^{\circ})$. As can be seen, the preferred axis derived from the five-parameter fit is consistent with that derived from the three-parameter fit.

Although the SNe Ia provide a strict constraint on the anisotropy of the universe, the redshifts of the SNe Ia in the Union2 dataset are no larger than 1.4. It is questionable whether such a constraint is suitable at high redshift region. The GRBs provide a good supplement to the shortcoming of SNe Ia. In fact, GRBs have been widely used to constrain the cosmological models \cite{Schaefer2007,Wei2010,Cai:2013lja,Velten:2013ej,Wei:2013xx}. A lot of methods have been proposed to calibrate GRBs, such as the scatter method \cite{Ghirlanda:2004fs}, the luminosity distance method \cite{Ghirlanda:2004fs}, and the Bayesian method \cite{Firmani:2005gs}. Wei has calibrated 109 long GRBs using the Amati relation \cite{Wei2010}. Among the 109 GRBs, 50 of then have redshift $z<1.4$, and the rest 59 have redshift $1.4<z\leqslant 8.1$. The calibrating procedure is as follows. Firstly, derive the distance modulus for the 50 low-redshift GRBs by using cubic interpolation from the 557 Union2 SNe Ia. Then, transform the distance modulus to the luminosity distance using Eq.(\ref{eq:distance modulus}). Thus, the isotropic equivalent energy $E_{\rm iso}$ can be derived from the relation
\begin{equation}\label{eq:iso-energy}
  E_{\rm iso}=4\pi d_L^2S(1+z)^{-1},
\end{equation}
if the flux $S$ and the redshift $z$ are known. Then use the best fit method to derive the parameters $(\lambda, b)$ in the Amati relation
\begin{equation}\label{eq:Amati relation}
  \log\frac{E_{\rm iso}}{{\rm erg}}=\lambda+b\log\frac{E_{p,i}}{300{\rm keV}},
\end{equation}
where $E_{p,i}=E_{p,{\rm obs}}\times(1+z)$ is the cosmological rest-frame spectral peak energy, and $E_{p,{\rm obs}}$ is the observed peak energy. Next, extend the parameters $(\lambda, b)$ to the rest 59 high-redshift GRBs and calculate their isotropic equivalent energy $E_{\rm iso}$ using the Amati relation Eq.(\ref{eq:Amati relation}). Finally, the luminosity distance and distance modulus of the high-redshift GRBs can be deduced from Eq.(\ref{eq:iso-energy}) and Eq.(\ref{eq:distance modulus}), respectively. As is showed, the advantage of this method is that it is cosmological model independent. The shortcoming is that it depends on the SNe Ia data.

\begin{figure}
  \centering
 \includegraphics[width=12 cm]{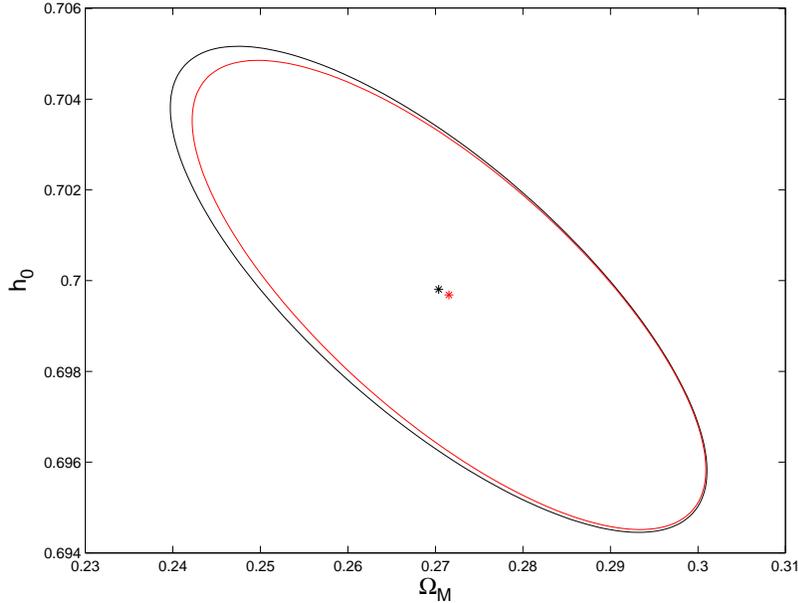}
 \caption{\small{The confidence regions of $1\sigma$ in the $(\Omega_M,h_0)$ plane. The black contour corresponds to SNe Ia only, the red contour corresponds to SNe Ia and GRBs. The center values are denoted by stars.}}
  \label{fig:error_ellipse_Mh}
\end{figure}

\begin{figure}
  \centering
 \includegraphics[width=12 cm]{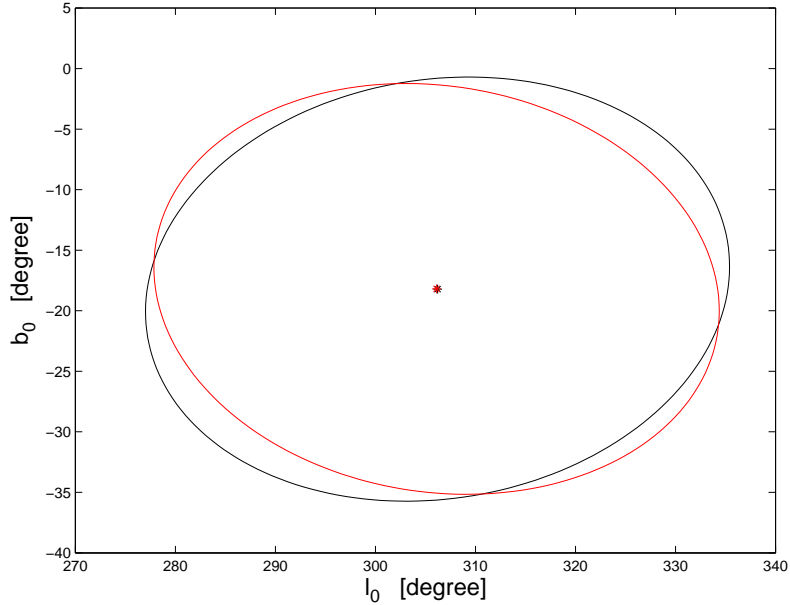}
 \caption{\small{The confidence regions of $1\sigma$ in the $(l_0,b_0)$ plane. The black contour corresponds to SNe Ia only, the red contour corresponds to SNe Ia and GRBs. The center values are denoted by stars.}}
  \label{fig:error_ellipse_lb}
\end{figure}

We combine the 59 high-redshift GRBs with the 557 SNe Ia in the Union2 dataset to constrain the parameters of the anisotropic cosmological model. Thus, our sample consists of 616 objects in total. Following the procedure above, we firstly apply two-parameter fit to the $\rm \Lambda$CDM model. It gives $(\Omega_M,h_0)=(0.272\pm 0.019, 0.700\pm 0.003)$. The $1\sigma$ confidence regions in the $(\Omega_M,h_0)$ plane are plotted in Figure \ref{fig:error_ellipse_Mh}. Then we fix $(\Omega_M,h_0)$ to their central values and apply three-parameter fit to derive the anisotropy of the universe, $D=-0.044\pm 0.018$, $(l_0,b_0)=(306.1^{\circ}\pm 18.7^{\circ}, -18.2^{\circ}\pm 11.2^{\circ})$. The $1\sigma$ confidence regions in the $(l_0,b_0)$ plane are depicted in Figure \ref{fig:error_ellipse_lb}. Finally, freeing all the five parameters, we get $(\Omega_M,h_0)=(0.269\pm 0.019,0.703\pm 0.004)$, $D=-0.050\pm 0.019$, and $(l_0,b_0)=(306.9^{\circ}\pm 16.1^{\circ}, -14.6^{\circ}\pm 10.6^{\circ})$. Note that all these results are consistent with that derived from the SNe Ia only.

\section{Discussions and conclusions}\label{sec:discussion}

Finsler geometry is a generalization of Riemann geometry \cite{Finsler:1951hf,Bao:2000jh}. Unlike Riemann geometry, Finsler geometry gets rid of the quadratic restriction on the line element and it is intrinsically anisotropic. The isometric transformation shows that there are less symmetries in Finsler geometry than in Riemann geometry \cite{Wang:1947kl,Rutz:1996pg,Li:2012km}. As a special kind of Finsler geometry, the Randers spacetime provides us an ideal framework to describe the anisotropic universe. By adding an arbitrary 1-form to the well-known FRW line element, a privileged axis in the universe is picked out. The distance-redshift relation is modified to be direction-dependent. The combination of the Union2 dataset with 59 long GRBs gives strict constraints on the anisotropic cosmological model. A direct fit to the dataset shows that the magnitude of anisotropy is $D=-0.044\pm 0.018$, and the privileged axis points towards $(l_0,b_0)=(306.1^{\circ}\pm 18.7^{\circ}, -18.2^{\circ}\pm 11.2^{\circ})$ in the GCS. In a previous paper \cite{Chang2013}, Chang et al. used the Union2 dataset to fit the model and got the privileged axis to be $(\alpha_0,\delta_0)=(263^{\circ}\pm 43^{\circ}, 91^{\circ}\pm 13^{\circ})$ in the ECS, and then transformed it into GCS as $(l_0,b_0)=(304^{\circ}\pm 43^{\circ}, -27^{\circ}\pm 13^{\circ})$. This result is somewhat misleading. The first thing is that, the declination of the ECS should be in the range $-90^{\circ}\leq \delta_0\leq 90^{\circ}$. The result $\delta_0=91^{\circ}$ is out of this boundary. We checked carefully the program and found a bug. The second thing is the error propagation between the two systems. The transformation from ECS to GCS is highly nonlinear. Thus, the error can't be simply copied from one system to the other. In order to avoid such a problem, we take the numerical analysis in the GCS directly. Although the best-fit parameters are close to that of previous paper, the statistical significance is highly improved.

\begin{figure}
  \centering
 \includegraphics[width=12 cm]{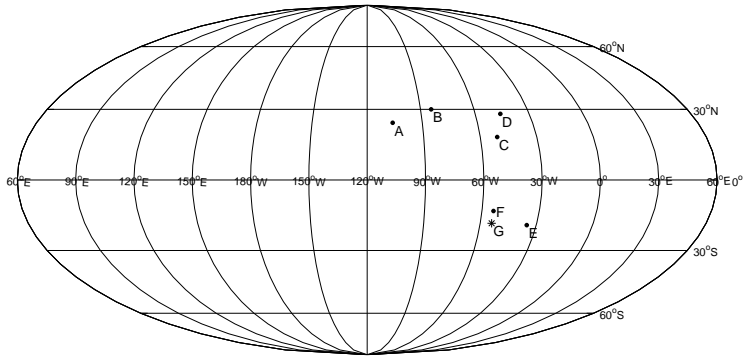}
 \caption{\small{The privileged axes in different publications in the GCS. A$(253.9^{\circ}, 24.1^{\circ})$ \cite{Jackson2012}, B$(276^{\circ}, 30^{\circ})$\cite{Kogut1993}, C$(309^{\circ}, 18^{\circ})$ \cite{Antoniou2010}, D$(314^{\circ}, 28^{\circ})$ \cite{CaiTuo2012}, E$(325^{\circ}, -19^{\circ})$ \cite{Kalus:2013zu}, F$(306^{\circ}, -13^{\circ})$ \cite{Cai:2013lja}, and G$(306.1^{\circ},-18.2^{\circ})$ is our result.}}
  \label{fig:galactic_coordinate}
\end{figure}

The anisotropy of the universe has been widely studied using different datasets and methods. Jackson investigated 468 ultra-compact radio sources and found the smallest value of $\Omega_M$ being towards $(l_0,b_0)=(253.9^{\circ}, 24.1^{\circ})$, while the maximum value being towards its opposite direction \cite{Jackson2012}. Kogut et al. found that the CMB dipole in the Local Group points to $(l_0,b_0)=(276^{\circ}\pm 3^{\circ}, 30^{\circ}\pm 3^{\circ})$ \cite{Kogut1993}. Antoniou \& Perivolaropoulos analyzed the Union2 dataset using the hemisphere comparison method in the context of ${\rm\Lambda}$CDM model, and found that the hemisphere of maximum accelerating expansion rate is in the direction $(l_0,b_0)=(309^{+23}_{-03}, 18^{+11}_{-10})^{\circ}$, which also corresponding to the minimum value of $\Omega_M$ \cite{Antoniou2010}. Using the same data and method, Cai \& Tuo took the deceleration parameter as the diagnostic to quantify the anisotropy level in the $\omega$CDM model, and found that the maximum accelerating expansion direction is $(l_0,b_0)=(314^{+20}_{-13}, 28^{+11}_{-33})^{\circ}$ \cite{CaiTuo2012}. Kalus et al. tested the isotropy of the expansion of the universe by estimating the hemispherical anisotropy of SNe Ia Hubble diagrams at redshifts $z < 0.2$, and found the highest expansion rate is towards $(l_0,b_0)=(325^{\circ}, -19^{\circ})$, while the Hubble anisotropy is about $\Delta H/H\sim 0.026$ \cite{Kalus:2013zu}.  Cai et al. constructed a direction-dependent dark energy model phenomenologically \cite{Cai:2013lja}. The constraints from the Union2 dataset and GRBs give the maximum anisotropic deviation direction $(l_0,b_0)=(306^{\circ}, -13^{\circ})$, while the magnitude of anisotropy is about $g_0=0.030^{+0.010}_{-0.030}$. The directions obtained by different authors are plotted in Figure \ref{fig:galactic_coordinate} for comparison. Our result (denoted by a star in Figure \ref{fig:galactic_coordinate}) is very close to the directions obtained in Ref. \refcite{Kalus:2013zu} and Ref. \refcite{Cai:2013lja}.

In our analysis, the best-fit parameters obtained from the Union2 dataset and GRBs are very close to that derived from the Union2 dataset only. This is in our expectation. The first reason is that the distance modulus of GRBs is calibrated from the Union2 dataset. The other reason is that the number of GRB (59) is much smaller than the number of SNe Ia (557). In order to make the conclusion more convincing, we may add more GRBs to the dataset and use a different calibrating  method which is independent of SNe Ia. Other dataset, such as ultra-compact radio sources and quasars, can also be used. However, the uncertainties of these objects are much larger than that of SNe Ia. SNe Ia are still the best standard candles at the present time.

Interestingly, the magnitude of anisotropy $(|D| =0.044\pm 0.018)$ obtained here is approximately equal to that of CMB dipole. The recent released Planck data show that the dipole magnitude of CMB temperature fluctuations is about $A=0.07\pm 0.01$, and the dipole direction points to $(l_0,b_0)=(218.9^{\circ}\pm 15.4^{\circ},-21.4^{\circ}\pm 15.1^{\circ})$ \cite{Ade:2013nlj}. This further conforms the previous WMAP result \cite{Bennett:2011fg}. The privileged axis found in this paper points towards a direction which is almost perpendicular to the direction of CMB dipole. The discrepancy between these two directions may be due to the different scales they involve. CMB is observed to the distance with redshift larger than 1000, while the furthest supernova studied in this paper has redshift $z\approx 1.4$. Even if GRBs were included, the highest redshift is no more than 8.1. In a recent paper \cite{Chang:2013vla}, Chang \& Wang proposed an inflation model of the very early universe in Randers spacetime, and obtained the primordial power spectrum of the scalar perturbation with direction dependence. This model is consistent with the hemispherical asymmetry of the CMB temperature fluctuations observed by WMAP and Planck satellites. These phenomena imply that anisotropy may be an intrinsic property of the spacetime.

\section*{Acknowledgements}
We are grateful to Y.-G. Jiang and M.-H. Li for useful discussion. This work is funded by the National Natural Science Fund of China under Grant No. 11075166 and No. 11375203.

\end{document}